\documentclass[unsortedadress,preprint
 ,secnumarabic%
,amssymb, amsmath,nobibnotes,prd]{revtex4}
\begin{document}
\title{Noncommutative Self-dual Gravity}

\author{H. Garc\'{\i}a-Compe\'an}
\email{compean@fis.cinvestav.mx}
\affiliation{Departamento de F\'{\i}sica,
Centro de Investigaci\'on y de Estudios Avanzados del IPN\\
P.O. Box 14-740, 07000 M\'exico D.F., M\'exico}
\author{O. Obreg\'on}
\email{octavio@ifug3.ugto.mx}
\affiliation{Department of Applied Mathematics and Theoretical Physics\\
Wilberforce Road,
Cambridge CB3 0WA, UK}
\altaffiliation[Permanent address:]{
Instituto de F\'{\i}sica de la Universidad de Guanajuato,\\
\vskip -1truecm
P.O. Box E-143, 37150 Le\'on Gto., M\'exico}
\author{C. Ram\'{\i}rez}
\email{cramirez@fcfm.buap.mx}
\affiliation{Instituto de F\'{\i}sica de la Universidad de Guanajuato,\\
P.O. Box E-143, 37150 Le\'on Gto., M\'exico}
\altaffiliation[Permanent address:]{
Facultad de Ciencias F\'{\i}sico Matem\'aticas,\\
\vskip -1truecm
Universidad Aut\'onoma de Puebla,
P.O. Box 1364, 72000 Puebla, M\'exico}
\author{M. Sabido}
\email{msabido@ifug3.ugto.mx}
\affiliation{Instituto de F\'{\i}sica de la Universidad de Guanajuato\\ P.O. Box E-143, 37150 Le\'on
Gto., M\'exico}

\date{\today}

\begin{abstract} Starting from a self-dual formulation of gravity, we
obtain a noncommutative theory of pure Einstein theory in four dimensions.
In order to do that, we use Seiberg-Witten map. It is shown that the
noncommutative torsion constraint is solved by the vanishing of
commutative torsion. Finally, the noncommutative corrections to the action
are computed up to second order.

\end{abstract}
\vskip -1truecm
\maketitle

\vskip -1.3truecm
\newpage

\setcounter{equation}{0}

\section{Introduction}

Nowadays, there are two main candidates for a quantum theory of the gravitational field: string theory
(M-theory)  and loop quantum gravity. From the description of the low energy excitations of open strings, in
the presence of a NS constant background $B-$field, a noncommutative effective low energy gauge action
\cite{connes,ref3} appears in a natural way. It is known, from $M$(atrix)-theory that, at low energies, the
coordinates of a gas of D0 branes are described by matrices, which give origin to virtual effects
\cite{joe,bfss}. Such effects give rise to supergravitational interaction in eleven dimensions. Thus, gravity
seems to arise from noncommutativity.

Along these lines, noncommutative gauge theory, as a continuous deformation of the usual theory, has
attracted a lot of attention. Although gravitation does not arise in the low energy limit of open
string theory as a gauge theory, some interesting effects of gravity processes (like the
graviton-graviton-D-brane scattering, in the presence of a constant $B$-field) can be computed
\cite{ardalan}. However, a deeper study of the deformations of pure gravitational theories is still
needed. Thus, the study of models of noncommutative gravity, independently of how they could arise
from string or M-theory, could be important. Such models could be obtained starting from those
formulations of gravitation which are based on a gauge principle. One of these formulations is the
one of self-dual gravity (for a review, see \cite{reviewsd,carlip}), from which the hamiltonian
Ashtekar's formulation \cite{ashtekar} can be obtained \cite{samuel,js}. The properties of this
formulation have allowed the search for quantum gravity in the framework of loop quantum gravity and
quantum geometry (for a review, see \cite{carlip}). In this context, it would be interesting to
explore noncommutative loop quantum gravity.

On the other hand, there are proposals for a noncommutative formulation of gravitation \cite{cham0}, motivated
by the understanding of the short distance behavior of the gravitational field \cite{connesbook}. Proposals
based on the recent developments are given in \cite{cham1,moffat,cham,cham2,zanon}. In particular, in
\cite{cham,cham2,zanon} a Seiberg-Witten map for the tetrad and the Lorentz connection is given,
where these fields are taken as components of a SO(4,1) connection in the first work, and of a U(2,2)
connection in the others. In these works a MacDowell-Mansouri (MM) type action is considered, invariant under
the subgroup U(1,1)$\times$U(1,1), and an excess of degrees of freedom, additional to the ones of the
commutative theory, is handled by means of constraints. For other recent proposals of noncommutative gravity,
see \cite{chandia}.

On the other hand, in \cite{wess5} it has been shown that noncommutative gauge theories, based on the
Seiberg-Witten map, for any commutative theory invariant under a gauge group $G$, can be constructed. The
resulting noncommutative theory can be seen as an effective theory, invariant under the noncommutative
enveloping algebra transformations, and also under the commutative transformations of $G$. This results from
the fact that the Seiberg-Witten map may be seen as a sort of gauge fixing, in which the degrees of freedom
added by noncommutativity to the fields and to the transformation parameters, are substituted by expressions
depending on the commutative fields, in such a way that in the commutative limit the original theory is
obtained. In this way, a minimal version of the noncommutative standard model with the gauge group
SU(3)$\times$SU(2)$\times$U(1) has been proposed \cite{wess6}.

Following these ideas, starting from a SL(2,{\bf C}), self-dual connection, in \cite{topologico} we have given
a formulation for quadratic noncommutative topological gravitation, which contains the SO(3,1) signature and
Euler topological invariants. In fact, the noncommutative signature can be straightforwardly obtained, but the
Euler invariant cannot, as it involves the same difficulty as the MM action, which contains a contraction with
the Levi-Civita tensor, instead of the SO(3,1) trace. However, both invariants can be combined into an
expression given by the signature with a SO(3,1) self-dual connection, which amounts to the SL(2,{\bf C})
signature.

In this paper, taking the same SL(2,{\bf C}) connection as in \cite{topologico}, considering the
Pleba\'nski formulation \cite{pleban}, we make a proposal for a noncommutative theory of gravity,
which is fully invariant under the noncommutative gauge transformations \cite{mempleb}. The
Pleba\'nski formulation is written as a SL(2,{\bf C}) topological BF formulation, given by the trace
of the two-form B, times the field strength \cite{gary}. The contact with Einstein gravitation is
done trough constraints on the B-field, which are solved by the square of the tetrad one-form
\cite{pleban}. This theory can be restated in terms of self-dual SO(3,1) fields, the connection and
the antisymmetric tensor B. After the identification of the B two-form with the tetrad one-form
squared, a variation of this action with respect to the connection gives the vanishing of the
torsion. The resulting action contains Einstein gravitation plus an imaginary term, which vanishes
due to the Bianchi identities. The noncommutative version is obtained at the level of the SL(2,{\bf
C}) theory, by the application of the Moyal product and the Seiberg-Witten map. In this way, a
highly nonlinear theory is obtained, which depends on the commutative SL(2,{\bf C}) fields. These
fields are then written in terms of the SO(3,1) self-dual connection and B-field, and the connection
is written in terms of the tetrads. The consistency of the last step is ensured by the fact that the
variation of the action with respect to the noncommutative SL(2,{\bf C}) connection, gives an
equation which is exactly solved by the vanishing of the commutative torsion.

The paper is organized as follows. In section 2 we briefly overview the Seiberg-Witten map and enveloping
algebra. In section 3 we state the commutative self-dual gravity theory from which the noncommutative one is
obtained. In section 4 the noncommutative theory is formulated and corrections are computed. Finally, section 5
contains our conclusions.

\vskip 1truecm

\section{Noncommutative Gauge Symmetry and the Seiberg-Witten Map}

In this section, a few conventions and properties of noncommutative spaces will be given for future reference.
For recent reviews see e.g. \cite{douglas}.

Noncommutative spaces can be understood as generalizations of the usual quantum mechanical commutation
relations, by the introduction of noncommutative coordinates $x^{\mu}$ satisfying,

\begin{equation}
[ x^{\mu},x^{\nu} ] =i\theta ^{\mu\nu},  \label{comm}
\end{equation}
where $x^{\mu}$ are linear operators acting on the Hilbert space $L^2({\bf R}^{n})$,
and $\theta ^{\mu\nu}=-\theta ^{\nu\mu}$ are real numbers. Given this linear operator algebra
${\cal A}$, the Weyl-Wigner-Moyal correspondence establishes an isomorphic
relation between it and the algebra of functions on ${\bf R}^{n}$, with an associative
and noncommutative star-product, the Moyal $\star-$product.
Thus, the Moyal algebra ${\cal A}_{\star} \equiv {\bf R}^{n}_{\star}$ is, under certain
conditions, equivalent to the Heisenberg algebra (\ref{comm}). The Moyal
product is given by
\begin{equation}
f(x)\star g(x)\equiv \left[ \exp \bigg(\frac{i}{2}\theta^{\mu\nu}{\frac{\partial
}{\partial \varepsilon^\mu}}{\frac{\partial }{\partial \eta^\nu}} \bigg) %
f(x+\varepsilon )g(x+\eta )\right] _{\varepsilon =\eta =0}.  \label{moyal}
\end{equation}
Under complex conjugation it satisfies
$\left( \overline{f\star g}\right) = \overline{g} \star \overline{f}$.

As far as we will be working with a nonabelian group, we must
include also matrix multiplication, so an $\ast$-product will be used as the
external product of matrix multiplication with $\star$-product. In this case
hermitian conjugation is given by $\left( f\ast g\right) ^{\dagger
}=g^{\dagger }\ast f^{\dagger }.$ Inside integrals this product has the cyclicity property
$Tr\int f_{1}\ast f_{2}\ast f_{3}\ast \cdots \ast f_{n} =
Tr\int f_{n}\ast f_{1}\ast f_{2}\ast f_{3}\ast \cdots \ast f_{n-1}$.
In particular $Tr\int f_{1}\ast f_{2}=Tr\int f_{1} f_{2}$.
From now on we will understand that
the multiplication of noncommutative quantities is given by this $*-$product.

Thus, to any expression containing space-time functions, a noncommutative expression can be associated by
substitution of the usual product by this $\ast-$product. However, this procedure has the well
known ambiguity of the ordering of the resulting expression, which could be fixed by physical
considerations. In particular, in the case of gauge theories, we wish to have a noncommutative
theory, invariant under a suitable generalization of the gauge transformations. This
generalization frequently is used to fix, to some extent, the ordering
ambiguities.

Let us consider a theory, invariant under the action of the Lie group $G$,
with gauge fields $A_\mu$, and matter fields $\Phi$ which transform under the adjoint
representation ${\bf ad}$,
\begin{eqnarray}
\delta_{\lambda }{A}_{\mu} &=&\partial _{\mu}{\lambda}+i\left[\lambda,{A}_{\mu}\right], \nonumber\\
\delta_{\lambda }{\Phi} &=&i\left[\lambda,\Phi\right],
\label{trafolambda}
\end{eqnarray}
where $\lambda=\lambda^iT_i$, and $T_i$ ($i=1, \dots , dim G$) are the
generators of the Lie algebra ${\cal G}$ of $G$, in the adjoint representation.
These transformations are generalized for the noncommutative connection \cite{ref3}, and for the
adjoint representation as,
\begin{eqnarray}
\delta_{\widehat{\lambda}}\widehat{A}_{\mu} &=&\partial _{\mu}\widehat{\lambda
}+i\left[\widehat\lambda\stackrel{\ast}{,}\widehat{A}_{\mu}\right],\label{trafoanc}\\
\delta_{\widehat{\lambda}}\widehat{\Phi} &=&
i\left[\widehat\lambda\stackrel{\ast}{,}\widehat{\Phi}\right].
\label{trafopsinc}
\end{eqnarray}
The commutators $\left[A\stackrel{\ast}{,}B\right]\equiv A\ast B-B\ast A$, have the correct
derivative properties when acting on products of noncommutative fields. Due to
noncommutativity, commutators like $\left[\widehat\lambda\stackrel{\ast}{,}\widehat{A}_{\mu}\right]$
take values in the enveloping algebra ${\cal U}({\cal G},{\bf ad})$
of the adjoint representation of ${\cal G}$. Therefore, $\widehat\lambda$
and the
gauge fields $\widehat{A}_\mu$ will also take values in this algebra.
In general,
for some representation ${\bf R}$, we will denote it by ${\cal U} ({\cal G},{\bf R})$ the section
of the enveloping
algebra ${\cal U}$ of ${\cal G}$, which corresponds to the representation ${\bf R}$.

Let us write for instance $\widehat\lambda=\widehat\lambda^I T_I$ and
$\widehat A=\widehat{A}^I T_I$, then,
\begin{equation}
\left[\widehat\lambda\stackrel{\ast}{,}\widehat{A}_{\mu}\right]=
\frac{1}{2}\left\{\widehat{\lambda}^{I}\stackrel{\ast}{,}\widehat{A}_{\mu}^{J}\right\} \left[
T_{I},T_{J}\right] +
\frac{1}{2}\left[ \widehat\lambda^{I}\stackrel{\ast}{,}\widehat{A}_{\mu}^{J}\right]
\left\{ T_{I},T_{J}\right\}.
\end{equation}
Thus all the products of the generators $T_I$ will be needed in order to close the algebra
${\cal U}({\cal G},{\bf ad})$.
Its structure can be obtained by successively computing the commutators and anticommutators
starting from the generators of ${\cal G}$, until it closes,
\begin{equation}
\left[ T_{I},T_{J}\right]=i{f_{IJ}}^KT_{K}, \ \ \ \ \left\{ T_{I},T_{J}\right\}
= {d_{IJ}}^KT_{K}.  \nonumber
\end{equation}

The field strength is \cite{ref3}
$\widehat{F}_{\mu\nu} =\partial _{\mu}\widehat{A}_{\nu}-
\partial _{\nu}\widehat{A}_\mu-i [\widehat{A}_{\mu}\stackrel{\ast}{,}\widehat{A}_{\nu}]$,
hence it also takes values in ${\cal U}({\cal G},{\bf ad})$. From (\ref{trafopsinc}) it turns out that,
\begin{equation}
\delta_{\lambda }\widehat{F}_{\mu\nu} =i\left( \widehat{\lambda }%
\ast \widehat{F}_{\mu\nu}-\widehat{F}_{\mu\nu}\ast\widehat{\lambda }\right).\label{trafoefenc}
\end{equation}

We see that these noncommutative transformation rules can be obtained from the commutative ones, just by replacing
the ordinary product of functions by the Moyal product, with a suitable
product ordering. This allows to construct in a simple way invariant quantities.

If we wish to have a continuous commutative limit, the noncommutative fields $\widehat\Phi$
must be power series expansions of the noncommutativity parameter $\theta$,
starting from the commutative ones,
\begin{equation}
\widehat{\Phi}=\Phi+\theta^{\mu\nu}\Phi^{(1)}_{\mu\nu}+
\theta^{\mu\nu}\theta^{\rho\sigma}\Phi^{(2)}_{\mu\nu\rho\sigma}+\cdots \ \ . \label{camposnc}
\end{equation}

Thus, in such an expansion, the noncommutative fields will have in general an infinity of
independent components. Moreover, the noncommutative gauge fields will take values in the enveloping algebra and, unless the enveloping algebra coincides with the Lie algebra of the
commutative theory, as is the case of $G=U(N)$, they will have also a bigger number of matrix
components. As this is also the case for the transformation parameters, it will be possible
to
eliminate a lot of degrees of freedom by fixing the gauge.

However, the Seiberg-Witten map \cite{ref3} establishes a one to one
correspondence among
the physical degrees of freedom of the noncommutative fields and the
physical degrees of freedom of the commutative fields. This fact is used in reference \cite{wess5}
to construct noncommutative gauge theories, in principle for any Lie group $G$.

The main point is that the Seiberg-Witten map allows for field dependent transformations. This means
that if we combine two transformations, the gauge parameters will be transformed as well.
Thus, if for an infinitesimal transformation matrix we have the correspondence
$\lambda\rightarrow\widehat\lambda$, to the commutator of two transformations will not correspond
simply the Moyal commutator, but \cite{wess5},
\begin{equation}
\widehat{[\lambda,\eta]}=[\widehat\lambda\stackrel{*}{,}\widehat\eta]+
i\left(\delta_\lambda\widehat\eta-\delta_\eta\widehat\lambda\right).\label{parametros}
\end{equation}

If we suppose that there is an expansion like (\ref{camposnc}) for these matrices,
\begin{equation}
\widehat{\lambda}=\lambda+\theta^{\mu\nu}\lambda^{(1)}_{\mu\nu}+
\theta^{\mu\nu}\theta^{\rho\sigma}\lambda^{(2)}_{\mu\nu\rho\sigma}+\cdots \ \ , \label{lambdanc}
\end{equation}
then a solution for the coefficients can be obtained \cite{ref3,wess5},
\begin{equation}
\widehat{\lambda }\left( \lambda ,A\right) =\lambda +\frac{1}{4}\theta ^{\mu\nu}\left\{ \partial
_{\mu}\lambda ,A_{\nu}\right\} +{\cal O}\left( \theta ^{2}\right). \label{difeq}
\end{equation}

Further, the Seiberg-Witten map determines the  $\Phi^{(a)}$ terms in (\ref{camposnc}), from the fact
that the noncommutative transformations are given by (\ref{trafoanc}), (\ref{trafopsinc})
and consequently, (\ref{trafoefenc}). These functions in (\ref{camposnc}) can be expressed in terms
of the commutative fields and their derivatives.
For the gauge fields, one solution is given by \cite{ref3},
\begin{equation}
\widehat{A}_{\mu}\left( A\right) =A_{\mu}-\frac{1}{4}\theta ^{\nu\rho}\left\{ A_{\nu},\partial
_{\rho}A_{\mu}+F_{\rho\mu}\right\} +{\cal O}\left( \theta ^{2}\right) ,  \label{asw}
\end{equation}
from which, for the field strength it turns out that,
\begin{equation}
\widehat{F}_{\mu\nu} =F_{\mu\nu}+\frac{1}{4}\theta ^{\rho\sigma}\bigg( 2\left\{
F_{\mu\rho},F_{\nu\sigma}\right\} -\left\{ A_{\rho},D_{\sigma}F_{\mu\nu}+\partial _{\sigma}F_{\mu\nu}\right\}
\bigg) +{\cal O}\left( \theta ^{2}\right). 
\label{swf}
\end{equation}

For fields in the adjoint representation we have got the solution,
\begin{equation}
\widehat{\Phi}\left(\Phi,A\right) =\Phi-\frac{1}{4}\theta^{\mu\nu}
\left\{A_\mu,(D_\nu+\partial_\nu)\Phi\right\}
 +{\cal O}\left( \theta ^{2}\right). \label{psisw}
\end{equation}

It is well known that these solutions are not unique, other terms depending on continuous
parameters can be added to them. In \cite{reno} this freedom has been related to
renormalizability properties. However, it can be also used in order to simplify the structure
of the theory \cite{wess6}. In particular, it allows to give the simple forms of (\ref{swf}) and
(\ref{psisw}), which have the interesting property that if the commutative fields vanish, also the
first order corrections will vanish. In this case, there is a solution for which all higher order
terms of the expansion (\ref{camposnc}) vanish as well.

These higher order terms can be obtained from the the Seiberg-Witten maps for which
$\frac{\partial}{\partial\theta^{\mu\nu}}\widehat\lambda(\theta)$ and
$\frac{\partial}{\partial\theta^{\mu\nu}}\widehat\Phi(\theta)$ are solutions, i.e. from the solutions of the equations which result from the corresponding gauge transformations, given
by the $\theta-$derivatives of (\ref{lambdanc}) and (\ref{trafopsinc}),
\begin{eqnarray}
\theta^{\mu\nu}\left[\delta_\lambda\frac{\partial}{\partial\theta^{\mu\nu}}\widehat{\eta}-
\delta_\eta\frac{\partial}{\partial\theta^{\mu\nu}}\widehat{\lambda}\right]=
i\theta^{\mu\nu}&\biggl(&\left[\frac{\partial}{\partial\theta^{\mu\nu}}\widehat{\lambda}
\stackrel{\ast}{,}\widehat\eta\right]+
\left[\widehat\lambda\stackrel{\ast}{,}
\frac{\partial}{\partial\theta^{\mu\nu}}\widehat{\eta}\right]-\nonumber\\
&&\ \ \frac{\partial}{\partial\theta^{\mu\nu}}\widehat{[\lambda,\eta]}+
\frac{i}{2}\left\{\partial_\mu\widehat\lambda\stackrel{\ast}{,}\partial_\nu\widehat\eta\right\}
\biggl),\label{horder1}
\end{eqnarray}
and
\begin{equation}
\theta^{\mu\nu}\delta_\lambda \frac{\partial}{\partial\theta^{\mu\nu}}\widehat\Phi=
i\theta^{\mu\nu}\left(\left[\frac{\partial}{\partial\theta^{\mu\nu}}\widehat\lambda
\stackrel{\ast}{,}\widehat\Phi\right]+
\left[\widehat\lambda
\stackrel{\ast}{,}\frac{\partial}{\partial\theta^{\mu\nu}}\widehat\Phi\right]+
\frac{i}{2}\left\{\partial_\mu\widehat\lambda\stackrel{\ast}{,}\partial_\nu\widehat\Phi\right\}
\right),\label{horder2}
\end{equation}
where the $\theta^{\mu\nu}$ factor is included in order to take into account of the antisymmetry in $\mu$ and
$\nu$.
A solution to this equation can be obtained \cite{ref3} from the first order solution
$\Phi^{(1)}_{\mu\nu}$. Indeed, from this first order term,
by substitution of the commutative fields by
the noncommutative ones, with multiplication given by the $\ast$-product, full Seiberg-Witten
mapped fields $\widehat{\lambda^{(1)}_{\mu\nu}}$ and $\widehat{\Phi^{(1)}_{\mu\nu}}$ can be constructed. Hence,
from (\ref{difeq}) and (\ref{psisw}), we have,
\begin{eqnarray}
\theta ^{\mu\nu}\widehat{\lambda^{(1)}}_{\mu\nu} &=&\frac{1}{4}\theta ^{\mu\nu}\left\{ \partial
_{\mu}\widehat\lambda\stackrel{\ast}{,}\widehat A_{\nu}\right\}, \label{difeq1}\\
\theta ^{\mu\nu}\widehat{\Phi^{(1)}}_{\mu\nu} &=&-\frac{1}{4}\theta^{\mu\nu}
\left\{\widehat{A}_\mu\stackrel{\ast}{,}(\widehat{D}_\nu+\partial_\nu)\widehat{\Phi}\right\}. \label{psisw1}
\end{eqnarray}
Now, after some algebra, taking into account the noncommutative gauge transformations (\ref{trafoanc}), (\ref{trafopsinc}) and
(\ref{parametros}), we get,
\begin{eqnarray}
\theta^{\mu\nu}\left[\delta_\lambda\widehat{\eta^{(1)}}_{\mu\nu}-
\delta_\eta\widehat{\lambda^{(1)}}_{\mu\nu}\right]&=&
i\theta^{\mu\nu}\left(\left[\widehat{\lambda^{(1)}}_{\mu\nu}\stackrel{\ast}{,}\widehat\eta\right]+
\left[\widehat\lambda\stackrel{\ast}{,}\widehat{\eta^{(1)}}_{\mu\nu}\right]-
\widehat{[\lambda,\eta]^{(1)}}_{\mu\nu}+
\frac{i}{2}\left\{\partial_\mu\widehat\lambda\stackrel{\ast}{,}\partial_\nu\widehat\eta\right\}
\right)\nonumber\\
\theta^{\mu\nu}\delta_\lambda\widehat{\Phi^{(1)}}_{\mu\nu} &=&i\theta^{\mu\nu}
\left(\left[\widehat{\lambda^{(1)}}_{\mu\nu}\stackrel{\ast}{,}\widehat\Phi\right]+
\left[\widehat\lambda\stackrel{\ast}{,}\widehat{\Phi^{(1)}}_{\mu\nu}\right]+
\frac{i}{2}\left\{\partial_\mu\widehat\lambda\stackrel{\ast}{,}\partial_\nu\widehat\Phi\right\}\right). \label{trafosigma1}
\end{eqnarray}
These equations give solutions for (\ref{horder1}) and (\ref{horder2}), if the following identifications are
made \cite{ref3},
\begin{eqnarray}
\frac{\partial}{\partial\theta^{\mu\nu}}\widehat\lambda&=&\widehat{\lambda^{(1)}}_{\mu\nu},
\label{horderl}\\
\frac{\partial}{\partial\theta^{\mu\nu}}\widehat\Phi&=&\widehat{\Phi^{(1)}}_{\mu\nu}, \label{horder}
\end{eqnarray}
which at $\theta=0$ are identically satisfied. In fact, (\ref{horder}) is more general, it is valid
also for the connection \cite{ref3}, and as well for any field transforming under a
linear representation. From it, together with (\ref{horderl}), by successive derivations with
respect to $\theta$, a solution for all higher terms of the Seiberg-Witten map can be computed.

Now, we see that if the first order term $\Phi^{(1)}$ vanishes, by construction also
$\widehat{\Phi^{(1)}}$ will vanish, and consequently in this case $\widehat\Phi=0$ is a
consistent solution for the Seiberg-Witten map.

The fact that the components of the noncommutativity parameter $\theta$ are constant,
has the important consequence that Lorentz covariance and general covariance under
diffeomorphisms of the underlying manifold, are spoiled. The answer that is usually given to this
question is that, at the scale where noncommutativity is relevant, it is possible that nature has
not the same symmetries like in the commutative limit.

\section{Description of Self-dual Gravity}

One of the main features of the tetrad formalism of the theory of gravitation \cite{utiyama}, is that
it introduces local Lorentz SO(3,1) transformations. In this case, the generalized Hilbert- Palatini
formulation is written as $\int e_a^{\ \mu}e_b^{\ \nu}R_{\mu\nu}^{\ \ ab}(\omega)d^4x$, where $e_a^{\
\mu}$ is the inverse tetrad, and $R_{\mu\nu}^{\ \ ab}(\omega)$ is the so(3,1) valued field strength.
The decomposition of the Lorentz group as SO(3,1)=SL(2,{\bf C})$\otimes$SL(2,{\bf C}), and the
geometrical structure of four dimensional space-time, makes it possible to formulate gravitation as a
complex theory, as in \cite{pleban,ashtekar}. These formulations take advantage of the properties of
the fundamental or spinorial representation of SL(2,{\bf C}), which allows a simple separation of the
action on the fields of both factors of SO(3,1), as shown in great detail in \cite{pleban}. All the
Lorentz Lie algebra valued quantities, in particular the connection and the field strength, decompose
into the self-dual and anti-self-dual parts, in the same way as the Lie algebra so(3,1)=s$\ell$(2,{\bf
C})$\oplus$s$\ell$(2,{\bf C}). However, Lorentz vectors, like the tetrad, transform under mixed
transformations of both factors and so this formulation cannot be written as a chiral SL(2,{\bf C})
theory. Various proposals in this direction have been made (for a review, see \cite{reviewsd}).  In a
early formulation, this problem has been solved by Pleba\'nski \cite{pleban}, where by means of a
constrained Lie algebra valued two-form $\Sigma$, the theory can be formulated as a chiral SL(2,{\bf
C}) invariant BF-theory, Tr$\int \Sigma\wedge R(\omega)$. In this formulation $\Sigma$ has two
SL(2,{\bf C}) spinorial indices, and it is symmetric on them $\Sigma^{AB}=\Sigma^{BA}$, as any such
s$\ell$(2,{\bf C}) valued quantity. The constraints are given by $\Sigma^{AB}\wedge
\Sigma^{CD}=\frac{1}{3}\delta_{(A}^C\delta_{B)}^D\Sigma^{EF}\wedge \Sigma_{EF}$ and, as shown in
\cite{pleban}, their solution implies the existence of a tetrad one-form, which squared gives the
two-form $\Sigma$. In the language of SO(3,1), this two-form is a second rank antisymmetric self-dual
two-form, $\Sigma^{+ab}=\Pi^{+ab}_{\ \ \ cd}\Sigma^{cd}$, where $\Pi^{+ab}_{\ \ \
cd}=\frac{1}{4}\left(\delta_{cd}^{ab}+i\epsilon^{ab}_{\ \ cd}\right)$. In this case, the constraints
can be recast into the equivalent form $\Sigma^{+ab}\wedge
\Sigma^{+cd}=\Pi^{+abcd}\Sigma^{+ef}\wedge \Sigma^{+}_{\ ef}$, with solution
$\Sigma^{ab}=2e^{a}\wedge e^{b}$.

For the purpose of the noncommutative formulation, we will consider self-dual gravity in a somewhat
different way as in the papers \cite{pleban,ashtekar}. In this section we will fix our notations and conventions.

Let us take the self-dual SO(3,1) BF action, defined on a $(3+1)$-dimensional pseudo-riemannian manifold
$(X, g_{\mu \nu})$,
\begin{equation}
I=i{\rm Tr}\int_X \Sigma^+\wedge R^+=i \int_X \varepsilon ^{\mu \nu \rho \sigma }
\Sigma_{\ \mu \nu }^{+ \ ab}R_{\rho \sigma ab}^{+}(\omega )d^4x,\label{accion}
\end{equation}
where
$R_{\rho \sigma ab}^+=\Pi _{ab}^{+cd}R_{\rho \sigma cd}$,
is the self-dual SO(3,1) field strength tensor.
This action can be rewritten as
\begin{equation}
I=\frac{1}{2}\int_X \varepsilon^{\mu\nu\rho\sigma}\left(i\Sigma_{\mu\nu}^{\ \ ab}R_{\rho \sigma ab}+
\frac{1}{2}\varepsilon_{abcd}\Sigma_{\mu\nu}^{\ \ ab}R_{\rho \sigma}^{\ \ cd}\right)d^4x.
\end{equation}
If now we take the solution of the constraints on $\Sigma$, which we now write as
\begin{equation}
\Sigma_{\mu\nu}^{\ \ ab}=e_\mu^{\ a}e_\nu^{\ b}-e_\mu^{\ b}e_\nu^{\ a},\label{sigma}
\end{equation}
then
\begin{equation}
I=\int_X (\det e R+i\varepsilon^{\mu\nu\rho\sigma}R_{\mu\nu\rho\sigma})d^4x.\label{accion1}
\end{equation}
The real and imaginary parts of this action must be variated independently
because the fields are real.
The first part represents Einstein action in the
Palatini formalism, from which, after variation of the Lorentz connection, a vanishing torsion
$T_{\mu\nu}^{\ \ a}=0$ turns out.
As a consequence, the second term vanishes due to Bianchi identities.

The action (\ref{accion}) can be written as
\begin{equation}
I=i\int_X \varepsilon ^{\mu \nu \rho \sigma }
\Sigma_{\mu \nu }^{+\ ab}R_{\rho \sigma ab}(\omega^{+})d^4x,\label{accion3}
\end{equation}
where
$R_{\mu \nu }^{\ \ ab}(\omega^+)=
\partial _{\mu }\omega _{\nu }^{+ab}-\partial _{\nu}\omega _{\mu }^{+ab}+
\omega _{\mu }^{+ac}\omega _{\nu c}^{+b}-\omega _{\nu}^{+ac}\omega _{\mu c}^{+b}$.
From the decomposition SO(3,1)=SL(2,{\bf C})$\times$SL(2,{\bf C}), it turns out that
$\omega _{\mu }^{\ i}=\omega _{\mu}^{+0i}$ is a SL(2,{\bf C}) connection. Further, if we take
into account self-duality, $\varepsilon _{cd}^{\ \ ab}\omega _{\mu }^{+cd}=2i\omega _{\mu }^{+ab}$,
we get $\omega _{\mu }^{+ij}=-i\varepsilon _{\ \,k}^{ij}\omega _{\mu }^{\ k}$.
Therefore,
\begin{eqnarray}
R_{\mu \nu }^{\ \ 0i}(\omega ^{+}) &=&\partial _{\mu }\omega _{\nu
}^{i}-\partial _{\nu }\omega _{\mu }^{i}+2i\varepsilon _{jk}^{i}\omega _{\mu
}^{j}\omega _{\nu c}^{k}=R_{\mu \nu }^{\ \ i}(\omega ) \\
R_{\mu \nu }^{\ \ ij}(\omega ^{+}) &=&\partial_\mu\omega_\nu^{+ij}-\partial_\nu\omega_\mu^{+ij}+
2(\omega_\mu^{\ i}\omega_\nu^{\ j}-\omega_\nu^{\ i}\omega_\mu^{\ j})=
-i{\varepsilon ^{ij}}_kR_{\mu \nu}^{\ \ k}(\omega ),
\end{eqnarray}
where ${R_{\mu\nu}}^i$ is the SL(2,{\bf C}) field strength.

Similarly, we define $\Sigma _{\mu \nu }^{\ \ i}=\Sigma _{\mu \nu }^{+\,0i}$, which transforms in
the SL(2,{\bf C}) adjoint representation. From it we get,
$\Sigma _{\mu \nu }^{+\,ij}=-i\varepsilon_{\ \,k}^{ij}\Sigma _{\mu \nu }^{\ \ k}$. Thus,
the action (\ref{accion3}) can be written as a SL(2,{\bf C}) $BF$-action
\begin{eqnarray}
I&=&i\int_X \varepsilon ^{\mu \nu \rho \sigma }\left[ \Sigma _{\mu \nu}^{+\,0i}
R_{\rho \sigma 0i}(\omega ^{+})+
\Sigma _{\mu \nu }^{+\,ij}R_{\rho\sigma ij}(\omega ^{+})\right]d^4x
=-4i\int_X \varepsilon ^{\mu \nu\rho \sigma }\Sigma _{\mu \nu }^{\ \ i}
R_{\rho \sigma i}(\omega )d^4x\nonumber.
\end{eqnarray}
Therefore, if we choose the algebra $s\ell(2,{\bf C})$ to satisfy $[T_i,T_j]=2i\varepsilon_{ij}^{\ \,k}T_k$
and $Tr(T_iT_j)=2\delta_{ij}$, we have that (\ref{accion}) can be rewritten as the self-dual action
\cite{pleban},
\begin{equation}
I=-2i{\rm Tr}\int_X\Sigma\wedge R,\label{accion4}
\end{equation}
which is SL(2,{\bf C}) invariant.

If the variation of this action with respect to the SL(2,{\bf C}) connection $\omega$
is set to zero, we get the equations
\begin{equation}
\Psi^{\mu\, i}=\varepsilon ^{\mu \nu \rho \sigma }D_{\nu }\Sigma _{\rho \sigma }^{\ \ i}=
\varepsilon ^{\mu \nu \rho \sigma }\left( \partial
_{\nu }\Sigma _{\rho \sigma }^{\ \ i}+2i\varepsilon _{\ jk}^{i}\omega _{\nu }^{\ j}\Sigma
_{\rho \sigma }^{\ \ k}\right) =0.
\end{equation}
Taking into account separately both real and imaginary parts, we get,
in terms of the SO(3,1) connection,
\begin{equation}
\varepsilon^{\mu\nu\rho\sigma}D_\nu\Sigma_{\rho\sigma}^{\ \
ab}=\varepsilon^{\mu\nu\rho\sigma}\left(\partial_\nu\Sigma_{\rho\sigma}^{\ \ ab}+
\omega_\nu^{\ ac}\Sigma_{\rho\sigma c}^{\ \ \ b}-\omega_\nu^{\ bc}\Sigma_{\rho\sigma c}^{\ \ \ a}
\right)=0,\label{constriccion}
\end{equation}
which after the identification (\ref{sigma}), can be written as
\begin{equation}
\varepsilon^{\mu\nu\rho\sigma}(\partial_\nu e_\rho^{\ a}e_\sigma^{\ b}-
\partial_\nu e_\rho^{\ b}e_\sigma^{\ a}
+\omega_\nu^{\ ac}e_{\rho c}e_\sigma^{\ b}-\omega_\nu^{\ bc}e_{\rho c}e_\sigma^{\ a})=
\varepsilon^{\mu\nu\rho\sigma}(T_{\nu\rho}^{\ \ a}e_\sigma^{\ b}-T_{\nu\rho}^{\ \ b}e_\sigma^{\ a})=0.
\end{equation}
From which the vanishing torsion condition once more turns out.

\section{The Noncommutative Action}

We start from the SL(2,{\bf C}) invariant action (\ref{accion4}). From it, the noncommutative action
can be obtained straightforwardly as
\begin{equation}
\widehat{I} =-i{\rm Tr}\int_X \widehat{\Sigma }\wedge \widehat{R}.\label{accionnc}
\end{equation}
This action is
invariant under the noncommutative SL(2,{\bf C}) transformations
\begin{eqnarray}
\delta_{\widehat{\lambda}}\widehat\omega_\mu&=&\partial_\mu\widehat\lambda+
i[\widehat\lambda\stackrel{\ast}{,}\widehat\omega_\mu],\label{trafoomega}\\
\delta_{\widehat{\lambda}}\widehat\Sigma_{\mu\nu}&=&i[\widehat\lambda\stackrel{\ast}{,}\widehat\Sigma_{\mu\nu}].
\label{trafosigma}
\end{eqnarray}

Actually, in order to obtain the noncommutative
generalization of the Einstein equation, we could consider the real part of (\ref{accionnc}),
\begin{equation}
\widehat{I_E}=-\frac{i}{2} {\rm Tr}\int_X\left[\widehat\Sigma\wedge\widehat R-
(\widehat\Sigma\wedge\widehat R)^\dagger\right],\label{accion5}
\end{equation}
which is also invariant under (\ref{trafoomega}) and (\ref{trafosigma}).

In order to obtain the corresponding to the torsion condition, a $\omega$ variation of
(\ref{accionnc}) must be done.
Although we are considering the commutative fields as the fundamental ones, the action is written
in terms of the noncommutative ones. Furthermore, the relation between the
commutative and the noncommutative physical degrees of freedom is one to one \cite{ref3}. So
the equivalent to the variation of the action with respect to $\omega$ will be the variation
with respect to $\widehat\omega$. Thus we write,
\begin{eqnarray}
\delta _{\widehat{\omega }} \widehat{I} &=&8i{\rm Tr}\int_X \varepsilon ^{\mu \nu \rho \sigma }
\left( \partial _{\rho }\widehat{\Sigma}_{\mu \nu }-i
[\widehat{\omega }_{\rho}\stackrel{\ast}{,} \widehat{\Sigma}_{\mu\nu}]
\right) \ast
\delta \widehat{\omega}_\sigma=0,\label{torsion}
\end{eqnarray}
from which we obtain the noncommutative version of (\ref{constriccion})
\begin{equation}
\widehat{\Psi^\mu}=
\varepsilon ^{\mu \nu \rho \sigma }\widehat D_\nu \widehat \Sigma_{\rho\sigma}=0.\label{varonc}
\end{equation}

These equations are covariant under the noncommutative transformations
(\ref{trafoomega}) and (\ref{trafosigma}),
which means that their Seiberg-Witten expansion should be similar to the one of a matter field in
the
adjoint representation (\ref{psisw}). In this case we have that, if the commutative field vanishes, also the first order term of the noncommutative one will vanish.
If this happens, as shown at the end of section 2, all the higher orders would
vanish as well.
Thus, we could expect that a solution to Eq. (\ref{varonc}) would be given by the solution
of the commutative equation $\Psi^\mu=0$, and this is indeed the case.
If we compute the first order of the Seiberg-Witten expansion of the covariant derivative of $\Sigma$, we get,
\begin{equation}
\widehat D_\mu\widehat\Sigma_{\nu\rho}=D_\mu\Sigma_{\nu\rho}-\frac{1}{4}\theta^{\zeta\tau}
\bigg(\left\{\omega_\zeta,(D_\tau+\partial_\tau)D_\mu\Sigma_{\nu\rho}\right\}+2\left\{R_{\mu\zeta},
D_\tau\Sigma_{\nu\rho}\right\}\bigg)+{\cal O}\left( \theta ^{2}\right),
\end{equation}
from which it turns out that,
\begin{equation}
\widehat\Psi^\mu=\Psi^\mu-\frac{1}{4}\theta^{\nu\rho}
\bigg(\left\{\omega_\nu,(D_\rho+\partial_\rho)\Psi^\mu\right\}-2\left\{R_{\nu\rho},
\Psi^\mu\right\}+2\delta_\nu^\mu\left\{R_{\tau\rho},\Psi^\tau\right\}\bigg)
+{\cal O}\left( \theta ^{2}\right).
\end{equation}
Hence the solutions of the commutative equations $\Psi^\mu=0$, will be as well solutions of
$\widehat\Psi^\mu=0$. Furthermore, $\Psi^\mu=0$ are equivalent to set the torsion equal to zero,
that is, after the substitution
${\Sigma_{\mu\nu}}^{ab}={e_\mu}^a {e_\nu}^b-{e_\nu}^a {e_\mu}^b$, their solution is given by,
\begin{equation}
\omega_\mu^{\ ab}=\frac{1}{2}e^{a\nu}e^{b\rho}\left[
e_{\mu c}(\partial_\nu e_\rho^{\ c}-\partial_\rho e_\nu^{\ c})-
e_{\nu c}(\partial_\rho e_\mu^{\ c}-\partial_\mu e_\rho^{\ c})-
e_{\rho c}(\partial_\mu e_\nu^{\ c}-\partial_\nu e_\mu^{\ c})\right].
\label{omegae}
\end{equation}

Therefore we can compute the corrections of the noncommutative action (\ref{accion5}) as follows. First we
write the Seiberg-Witten expansion of the SL(2,{\bf C}) fields $\widehat\Sigma$ and $\widehat\omega$.
Furthermore the commutative SL(2,{\bf C}) fields are written by means of the self-dual SO(3,1) fields, $\omega
_{\mu }^{\ i}=\omega _{\mu}^{+0i}$ and $\Sigma _{\mu \nu }^{\ \ i}=\Sigma _{\mu \nu }^{+\,0i}$. Then, decompose
these self-dual fields into the real ones $\omega_\mu^{\ ab}$ and $\Sigma_{\mu\nu}^{\ \ ab}$, and then
substitute ${\Sigma_{\mu\nu}}^{ab}={e_\mu}^a {e_\nu}^b-{e_\nu}^a {e_\mu}^b$ and write the connection as in
(\ref{omegae}). In this case we will have a noncommutative action, which will depend only on the tetrad.

If we consider the real part, the first order correction vanishes, and, after a lengthy calculation, the second order one turns out to be, already written in terms of commutative SO(3,1) fields,
\begin{eqnarray}
&&\widehat I_{\theta ^{2}} =\frac{\theta ^{lm}\theta ^{rs}}{2^{4}}\int_X dx^{4}\det
(e)\bigg\{\partial _{s}\left[ \partial _{m}\left( \omega _{l}^{ab}R_{abcd}\right)
\omega _{r}^{cd}\right] +\partial _{s}\left( \omega _{l}^{ab}\partial
_{r}\omega _{m}^{cd}R_{abcd}\right) \nonumber \\
&& + 16R_{m}^{a}\left( R_{ar}^{bc} R_{lsbc}-\omega _{r}^{bc}\partial _{s}R_{albc}\right)
-\frac{1}{2}\partial _{m}\left( \omega_{r}^{ab}\left( \partial _{s}\omega
_{lab}+R_{slab}\right) \right) R-\frac{1}{2}\partial _{m}\omega
_{l}^{ab}(R_{ar}^{ef}R_{bsef}-\omega _{r}^{ef}\partial _{s}R_{abef})\nonumber \\
&&-2\partial _{r}\omega_{m}^{ef}\partial _{s}R_{ef}^{cd}\omega _{lcd}-
\frac{1}{2}\varepsilon _{abcd}\varepsilon ^{efgh}\bigg[\omega_{lef}\partial
_{m}\left( R_{gr}^{ab}R_{hs}^{cd}-\omega _{r}^{ab}\partial
_{s}R_{gh}^{cd}\right) +\frac{1}{4}\partial _{s}\left( \omega _{lef}\partial
_{r}\omega _{m}^{ab}\right) R_{gh}^{cd} \nonumber\\
&& + \partial_{m}\omega_{lef}\left( R_{gr}^{ab}R_{hs}^{cd}-\omega
_{r}^{ab}\partial _{s}R_{gh}^{cd}\right) -\frac{1}{2}\partial _{s}\left(
\partial _{m}\left( \omega _{l}^{ab}R_{gh}^{cd}\right) \omega
_{ref}\right)\bigg]\bigg\},
\end{eqnarray}
where the connection $\omega_\mu^{\ ab}$ is given by (\ref{omegae}). This correction terms could allow the
explicit computation of deformed known gravitational metrics. 

\section{Conclusions}

In this work we propose an ansatz to get a noncommutative formulation of standard four-dimensional Einstein
gravitation. We start from a self-dual SO(3,1) BF-action, which is equivalent to Einstein gravitation after the
substitution of the B-field in terms of the tetrad. This action is reformulated as the chiral SL(2,{\bf C})
invariant self-dual action (\ref{accion4}), from which the noncommutative action (\ref{accionnc}) is
straightforwardly obtained. This chiral action allows us to find an alternative noncommutative gravity action
in four dimensions, that generalizes the usual general relativity. As mentioned, there are other proposals
already introduced in the literature \cite{moffat,cham,cham2}. In our proposed action the noncommutative spin
connection variation gives the noncommutative `torsion condition' (\ref{varonc}), which is exactly solved by
(\ref{omegae}). This allows to introduce the tetrad at the commutative level in a consistent way (\ref{sigma}).
As a consequence, the corrections to Einstein gravitation can be, modulo a laborious algebra, straightforwardly
computed.

In the process we have used the results from \cite{wess5,wess6}, developed there to construct the
noncommutative versions of standard model and GUT's. In the present paper, the Seiberg-Witten map for matter
fields in the adjoint representation of any gauge group has been constructed in order to get the Seiberg-Witten
map for the $\Sigma$-field.

The physical consequences of the our noncommutative extension of standard general relativity remain to be
studied. An interesting possibility seems to be the study of inflation in this model. It is well known, in a
very different physical setting, that the trace anomaly that leads to higher derivative corrections in the
corresponding effective action, could produce inflation \cite{staro,inflation}.

Finally, the results presented here, can be regarded as a preliminary step for the construction of a
noncommutative version of Ashtekar´s hamiltonian formulation through a noncommutative Legendre transformation.
Moreover, the computation of noncommutative gravitational effects can be done, for instance, from the
corrections to known metrics.  The details of these works will be reported elsewhere. Eventually, it
would be interesting to search for a quantization of the results obtained in the present paper and
proceed to find the corresponding loop quantum
gravity.


\vskip 2truecm
\centerline{\bf Acknowledgments}
This work was supported in part by CONACyT M\'exico Grants Nos. 37851E and
33951E, as well as by the sabbatical grants 020291 (C.R.) and 020331 (O.O.).


\vskip 2truecm



\end{document}